\documentstyle[12pt,a4wide]{article}
\begin{document}
\newcommand{\ds}{\displaystyle}

\newcommand{\zz}{\cal Z}
\newcommand{\uz}{ {\bf z} }
\newcommand{\UZ}{ {\bf Z} }
\newcommand{\pr}{\prime}
\newcommand{\rr}{{\cal R}}
\newcommand{\uzb}{ \bar{\uz} }

\newcommand{\th}{\theta}
\newcommand{\tb}{\bar{\theta}}
\newcommand{\TH}{\Theta}
\newcommand{\TB}{\bar{\Theta}}
\newcommand{\la}{\lambda}
\newcommand{\LA}{\Lambda}
\newcommand{\LB}{\bar{\LA}}
\newcommand{\OM}{\Omega}
\newcommand{\DE}{\Delta}
\newcommand{\bt}{\bar{\tau}}
\newcommand{\zb}{\bar z}
\newcommand{\pa}{\partial}
\newcommand{\dab}{\bar D}
\newcommand{\ZB}{\bar Z}
\newcommand{\pab}{ \bar{\partial} }
\newcommand{\HT}{ {H_{\th}}^z }
\newcommand{\HB}{ {H_{\tb}}^z }
\newcommand{\HO}{ H_{\th} ^{\ \th} }
\newcommand{\HZ}{ H_{\zb} ^{\ \th} }
\newcommand{\HZB}{ H_{\zb} ^{\ z} }
\newcommand{\HOB}{ H_{\tb} ^{\ \th} }

\newcommand{\zt}{\tilde z}
\newcommand{\dt}{\tilde D}
\newcommand{\pabt}{\tilde{\pab}}

\newcommand{\thm}{\theta^-}
\newcommand{\tbm}{{\bar{\theta}}^-}

\newcommand{\lb}{\bar{\la}}
\newcommand{\kb}{\bar k}
\newcommand{\al}{\alpha}
\newcommand{\ab}{\bar{\al}}
\newcommand{\be}{\beta}
\newcommand{\bb}{\bar{\be}}
\newcommand{\ga}{\gamma}
\newcommand{\gb}{\bar{\ga}}
\newcommand{\de}{\delta}
\newcommand{\db}{ \bar{\de}}
\newcommand{\qb}{{\bar Q}}
\newcommand{\mb}{{\bar{\mu} }}

\newcommand{\nab}{\nabla}
\newcommand{\nabar}{\bar{\nabla}}

\newcommand{\pal}{{\rm pal}}

\newcommand{\vv}{{\cal V}}

\newtheorem{definition}{Definition}
\newtheorem{theorem}{Theorem}


\thispagestyle{empty}


\begin{center}
{\bf \Huge{$N=2$ supersymmetric}}
\end{center}
\begin{center}
{\bf \Huge{pseudodifferential symbols}}
\end{center}
\begin{center}
{\bf \Huge{and super {\cal W}-algebras}}
\end{center}

\vskip 0.9truecm
\centerline{
{\bf St\'ephane Gourmelen}}
\bigskip
\bigskip
\centerline{{\it Institut de Physique Nucl\'eaire de Lyon,
IN2P3/CNRS}}
\centerline{\it Universit\'e Claude Bernard}
\centerline{\it 43, boulevard du 11 novembre 1918}
\centerline{\it F - 69622 - Villeurbanne Cedex}

\vskip 0.9truecm

\nopagebreak
\begin{abstract}
We study the superconformally covariant 
pseudodifferential symbols defined on 
$N=2$ super Riemann surfaces. This allows us to construct a 
primary basis for $N=2$ super ${\cal W}_{KP}^{(n)}$-algebras 
and, by reduction, for $N=2$ super ${\cal W}_{n}$-algebras.
\end{abstract}


 \section{Introduction}

${\cal W}$-symmetry plays an important r\^ole 
in the context of two-dimensional 
conformal field theories \cite{bouschou,ketov} 
and their applications to  
critical phenomena and to string theory \cite{dfgzj}.
Classical ${\cal W}$-algebras first appeared in the study of
integrable systems, namely of generalized KdV hierarchies \cite{dick}.
More precisely, the ${\cal W}_n$-algebra arises 
as the second hamiltonian structure 
of the $n$-th KdV hierarchy whose   
Poisson brackets are defined on the manifold 
of differential operators of order $n \geq 2$.
In the simplest case ($n=2$), this algebra coincides 
with the Virasoro algebra, as was first noticed in ref. \cite{gerneu}. 
This fact exhibits the connection between ${\cal W}_n$-algebras 
and conformal field theory and suggests to apply  
conformal symmetry to the formulation of classical ${\cal W}_n$-algebras.
This was done in ref. \cite{dfiz.1} (see also references therein) where 
it was shown that 
the ${\cal W}_n$-algebra possesses a `primary basis' of generators
 consisting of a projective connection (the Virasoro generator) 
and $n-2$ primary fields which transform like $k$-forms ($k=3,..,n$) 
under the Virasoro flow.

Besides 
${\cal W}_n$-algebras there are other ${\cal W}$-algebras
 which are said to be `infinite' 
 because they contain  an infinity of independent generators.
These algebras are related to each other by 
reductions, truncations or contractions \cite{fmrtopo}. 
In particular, every ${\cal W}_n$-algebra can be obtained by reduction 
from the infinite ${\cal W}_{KP}^{(n)}$-algebra \cite{fmrkpq}. 
The latter is obtained by applying the second 
hamiltonian structure of ${\cal W} _n$ to 
pseudodifferential symbols rather than differential operators.
It was shown in ref. \cite{hua1} that such a symbol can be parametrized 
by a projective connection and an infinity of primary 
fields, thus providing  
a primary basis for every 
${\cal W}_{KP}^{(n)}$-algebra ($n \geq 2$).

${\cal W}$-algebras admit supersymmetric extensions which 
manifest themselves in the context of superstring 
or super-Toda field theories.
The second hamiltonian structure of $N=1$ super ${\cal W}$-algebras 
was constructed in superspace in ref. \cite{frtsusy1}. 
Their primary basis was determined in ref. \cite{gt1} for 
super ${\cal W}_n$ and generalized to super ${\cal W}_{KP}^{(n)}$ \cite{hua2}.
 In this paper, 
we are interested in $N=2$ super ${\cal W}$-algebras. They have been 
extensively studied in the $N=1$ formalism (see \cite{frtsusy2,hn,gt2,hua3} 
and references therein) until their formulation in $N=2$ superspace 
was discovered \cite{pop.2,dg}. $N=2$ super ${\cal W}_n$-algebras 
were shown to admit a primary basis which was constructed in ref. \cite{gg}.
The aim of this paper is to study the $N=2$ supersymmetric pseudodifferential 
symbols and to apply them to the determination of a primary basis 
for $N=2$ super 
${\cal W}_{KP}$-algebras in $N=2$ superspace.

In section 2, we summarize some concepts and tools of $N=2$ 
superconformal symmetry and superconformally covariant operators.
These are used in section 3 to study $N=2$ supersymmetric 
pseudodifferential symbols. We pay a particular attention to 
the so-called Bol symbols (parametrized by a superprojective connection) 
which are studied systematically. A particular class 
of them is generalized in section 4 and applied to the 
formulation of $N=2$ super ${\cal W}_{KP}$-algebras in a primary basis.


\section{$N=2$ supersymmetric differential operators}
For further details concerning the notions 
summarized in this section, we refer to ref. \cite{gg}.

\subsection{Geometric framework}
   \label{geofra}
   
\paragraph{$N=2$ supersymmetry}

In
order to make $N=2$ supersymmetry manifest, 
all considerations will be carried out on a compact two-dimensional $N=2$
supermanifold $\Sigma$  \cite{dewitt} with local coordinates $\uz \equiv
(z, \th, \tb)$ and their complex conjugates ({\em c.c.}) $\bar{\uz}  \equiv
(\zb , \thm, \tbm )$.  Here, $z, \zb$ are even and $\th, \tb, \thm, \tbm$ are odd
Grassmann numbers. The tangent space is spanned by the derivatives $( \pa ,
\, D  , \,
\dab)$ (and their {\em c.c.} $( \pab ,
\, D_-  , \,
\dab _-)$) defined by
$\pa =\ds \frac{\pa}{\pa z}
\  , \ 
D=\ds \frac{\pa}{\pa \th}  +  \frac{1}{2} \, \tb \pa
\ , \ \dab = \ds \frac{\pa}{\pa \tb}  +  \frac{1}{2} \, \th  \pa
\ \ .$
Their graded Lie brackets
$\{  D ,  \dab \} =  \pa
\  , \ 
D^2   =  0  = \dab^2$
are those of the $N=2$ supersymmetry algebra.

\paragraph{$N=2$ superconformal symmetry}

Since we will  be interested in $N=2$
superconformal symmetry, we require the supermanifold $\Sigma$ to be a
super Riemann Surface (SRS). This means that local
coordinate systems on $\Sigma$ are related by superconformal
transformations.
By definition \cite{dgg}, a {\em superconformal transformation} of local coordinates
of $\Sigma$ is a superdiffeomorphism $({\uz} ; {\uzb} ) 
\longmapsto
({\uz}^{\prime} ; {\uzb}^{\prime})$ satisfying the following three 
properties (as well as the {\em c.c.} relations): 
\begin{eqnarray*}
{\rm (i)} & \quad &
{\uz}^{\prime}  =  {\uz} ^{\prime} ( {\uz} )
\quad \Longleftrightarrow \quad
D_- {\uz}^{\prime}  = 0 =
\dab_- {\uz}^{\prime}
\nonumber
\\
\label{3i}
{\rm (ii)} & \quad &
D \tb^{\prime} \ = \ 0 \ =
\dab \th^{\prime}
\\
{\rm (iii)} & \quad &
D z^{\prime}  =
\frac{1}{2}
\tb ^{\prime} \, D \th ^{\prime}
\quad \qquad , \quad \
\dab z^{\prime} \ = \ \frac{1}{2}
\th ^{\prime} \, \dab \tb ^{\prime}
\nonumber
 \ \ \ .
\end{eqnarray*}
These relations imply
that $D$ and $\dab$ transform homogeneously,
\begin{displaymath}
   \label{7a}
D ^{\prime} \  = \  {\rm e} ^w \ D
\quad , \quad
\dab ^{\prime} \  = \  {\rm e} ^{\bar{w}} \ \dab
\ \ ,
\end{displaymath}
where
$ \ {\rm e}^{-w}  \equiv  D \th^{\prime}
 \ , \ D w \ = \ 0$ and 
${\rm e}^{-\bar{w}}  \equiv  \dab  \tb^{\prime}
\ ,\ 
\dab  \bar{w} \ = \ 0$.

\paragraph{Superprojective connection}

The {\em super Schwarzian derivative} 
${\cal S} ({\uz} ^{\prime} , {\uz})$ 
associated to a superconformal change
of coordinates $\uz \longmapsto \uz^{\pr} (\uz )$ is defined by 
\begin{eqnarray*}
-{\cal S} ({\uz} ^{\prime} , {\uz} )
& = & 
\frac{\pa \dab \tb^{\prime}}{\dab \tb^{\prime}}
\ - \
\frac{\pa D\th^{\prime}}{D \th^{\prime}}
\ + \
\frac{ \pa \tb^{\prime} }{ \dab \tb^{\prime} }
\; \frac{ \pa \th^{\prime} }{ D \th^{\prime} }
\nonumber  \\
& = &
2 \, {\rm e}^{-{1 \over 2} (w + \bar w)} \
[D,\dab]
\, {\rm e}^{{1 \over 2} (w + \bar w)}
\ \ .
\end{eqnarray*}

The following study will make an extensive use of a {\em
superprojective connection}. This is a superfield 
${\cal R} \equiv {\cal R}_{\th \tb} (\uz )$ which is   
locally superanalytic ({\em i.e.} $D_- {\cal R} =0= \dab_- {\cal R}$) and 
which transforms
under a superconformal transformation of coordinates 
according to
\begin{displaymath}
{\cal R}^{\prime}
({\uz} ^{\prime})
= {\rm e}^{w+\bar w} \left[ {\cal R} ({\uz})
- {\cal S} ({\uz} ^{\prime} , {\uz} )
\right]
\ \ .
\end{displaymath}
Such a field
can be globally defined on compact SRS's of arbitrary
genus \cite{cco,dgg}.

\subsection{Superconformally covariant differential operators}
\label{diffop}
A superconformal (or primary) field of superconformal
weight $(p,q)$ is a function $C_{p,q} \in C ^{\infty} (\Sigma)$ ({\em i.e.} the space
of supersmooth functions on $\Sigma$ \cite{dewitt}) which transforms under a
superconformal change of local coordinates according to 
\begin{equation}
\label{tra}
C_{p,q}^{\prime} ({\uz}^{\prime} ; {\uzb}^{\prime}) = {\rm e}^{pw+q \bar w}
C_{p,q} ({\uz} ; {\uzb})\qquad  ( \, p,q \in {\bf Z}/2
\ \ , \ \ p+q \in {\bf Z} \, )
\ \ \ .
\end{equation}
The space of these fields will be denoted by ${\cal F}_{p,q}$.
\begin{definition}
   \label{def1}
A generic {\em superdifferential operator} on $\Sigma$
is locally
defined by
\begin{equation}
{\cal L} \ = \ \sum_{n=0}^{n_{max}} \ \left(\ a_n \ + \ \alpha_n D \
+\
\beta_n \dab \ +\ b_n [ D,\dab ] \right) \pa ^n \ \ \ ,
\label{genopdiff}
\end{equation}
where $a_n,b_n$ and $\alpha_n,\beta_n$ are, respectively, even and odd
superfields belonging to $C ^{\infty} (\Sigma)$. Such an operator is 
called {\em superconformally covariant} (or {\em covariant} for short) if it maps
primary fields of some weight $(p,q)$ to primary fields of some weight
$(p^{\prime},q^{\prime})$:
\begin{equation}
   {\cal L} \equiv {\cal L}_{p,q} \ :\ 
{\cal F}_{p,q} \ \longrightarrow
{\cal F}_{p^{\prime},q^{\prime}} \ \ .
\label{eqdef1}
\end{equation}
\end{definition}
Note that the correspondence (\ref{eqdef1}) is equivalent to the following 
transformation law under superconformal changes of local coordinates :
\begin{equation}
\label{translaw}
      \left( {\cal L} _{p,q} \right) ^{\prime} \ =\ 
      {\rm e}^{p ^{\prime}w+q ^{\prime}\bar w}
      \, {\cal L} _{p,q} \,
      {\rm e}^{-p w-q \bar w} \ \ .
      \end{equation}

In order to construct such covariant operators, it is
convenient to use a {\em superaffine connection}. This is a collection of
superfields $B \equiv B_{\th} ,\bar B \equiv \bar B _{\tb}$ which
are locally defined on $\Sigma$ and which satisfy the following three 
conditions :
they\ are\ locally\ superanalytic, 
they satisfy the chirality conditions
$D\, B = 0 = \dab \, \bar B$
and they 
transform
under a superconformal change of local coordinates
 according to 
$B^{\prime}
({\uz} ^{\prime} )
= {\rm e}^{w} \left[ B ({\uz} )
+ D  \bar w \right] \ \ ,\ \ 
\bar B ^{\prime}
({\uz} ^{\prime} )
= {\rm e}^{\bar w} \left[ \bar B ({\uz} )
+ \dab    w \right]$.

Using an affine connection one can introduce
{\em supercovariant derivatives}
\begin{eqnarray*}
 \label{cd}
 \nabla \equiv \nabla_{p,q}  =  D - q B &:&
 {\cal F}_{p,q} \
\longrightarrow {\cal F}_{p+1,q} \\
 \bar{\nabla} \equiv \bar{\nabla}_{p,q}  =  \dab  - p
 \bar B &:&
 {\cal F}_{p,q} \ \longrightarrow
{\cal F}_{p,q+1} \ \ .
\end{eqnarray*}
By construction, these  
are nilpotent: 
   $\nabla ^2 \ =\  0\ =\ \bar{\nabla} ^2$.

The most general covariant operator that is locally defined 
on $\Sigma$ is simply obtained
by replacing the two fermionic derivatives $D$ and $\dab$ in expression
(\ref{genopdiff}) by supercovariant ones, $\nabla$ and  $\bar{\nabla}$:
\begin{equation}
{\cal L}_{p,q} \ = \ \sum_{n=0}^{n_{max}} \ \left(\ a_n \ + \ \alpha_n
\nabla \
+\
\beta_n \bar{\nabla} \ +\ b_n [ \nabla ,\bar{\nabla} ] \right) \{
\nabla ,\bar{\nabla} \}^n 
\label{genopdiffcov} \ \ \ .
\end{equation}

\subsection{Bol differential operators}
   \label{bolop}

The only compact SRS's which admit a globally defined affine
connection are those of genus one
\cite{cco,dgg}. 
On the other hand, 
affine and projective connections are locally related by
the {\em super Miura
transformation}
\footnote{
In order to avoid ambiguities, 
 we adopt from now on the following notation for the action of derivatives
 on a field $C$: $(\pa C),\, (D C),\, (\dab C), ([ D, \dab ] C),$... 
 denote derivatives of the field $C$ 
while a derivative acts operatorially otherwise, e.g. 
$\pa C= (\pa C)\, +\, C \pa$.}

\begin{equation}
\label{miu}
{\cal R}
 \ = \  (D \, \bar B )-  (\dab \, B)  - B \, \bar B
\ \ .
\end{equation}
The fact that a projective connection can
always be defined globally on a SRS motivates the following definition.
\begin{definition}
   \label{def2}
A {\em Bol 
operator} is a covariant differential 
operator on $\Sigma$ which depends on an unique superfield, namely a projective
connection.
\end{definition}

It follows directly from this definition that a Bol operator is globally
defined on the SRS. In order to construct a Bol operator, 
we require that the operator
${\cal L} _{p,q}$ of eq. (\ref{genopdiffcov}) only depends on the affine
connections $B, \bar{B}$ through the combination $\rr = D\bar B - \dab B -
B\bar B$ given by the Miura transformation. By
using a variational argument 
(one imposes that $\delta {\cal L}_{p,q} =0$ while varying $B$ and $\bar B$
subject to the condition that $\rr$ is fixed), one is led to the following
 result \cite{gg} :
\begin{theorem}
For each superconformal weight $(p,q) \in ({\bf Z}/2, {\bf Z}/2)$
such that $-(p+q) \in {\bf N}^{*}$, there exists a Bol
operator defined on the whole space ${\cal F}_{p,q}$.  This operator
is unique up to a global factor and 
is of order $n=-(p+q)$. It reads
\begin{equation}
L_{p,q} ({\cal R}) \ = \ q (\nabla \bar{\nabla})^n
-p (\bar{\nabla} \nabla )^n 
\ : \ {\cal F}_{p,q} \ \longrightarrow
{\cal F}_{p+n,q+n} \ \ .
\label{resbol}
\end{equation}
\end{theorem}
Note that there are other Bol operators which are only defined on appropriate
subspaces of ${\cal F}_{p,q}$ \cite{gg}. 

\section{$N=2$ supersymmetric pseudodifferential symbols}

\subsection{Basic definitions and relations}

We aim to extend the analysis of sections  \ref{diffop} and
\ref{bolop} to the pseudodifferential case.
A generic {\em pseudodifferential symbol} (or {\em symbol} for short) is locally
defined on $\Sigma$ by \cite{pop.2}
\begin{equation}
{\cal L} \ = \ \sum_{n=-\infty}^{n_{max}} \ \left(\ a_n \ + \ \alpha_n D \
+\
\beta_n \dab \ +\ b_n [ D,\dab ] \right) \pa ^n
\label{genpdo} \ \ ,
\end{equation}
where we have introduced the inverse $\pa ^{-1}$ of the usual derivative :
\begin{equation}
   \pa \, \pa ^{-1} \ =\ \pa ^{-1} \, \pa \ =\ 1 \ \ .
   \label{pa-1}
\end{equation}
The symbol ${\cal L}$  can be divided
into its {\em differential part} (the summation going from $n=0$ to $n_{max}$) and
its {\em integral part} (the summation going from $n=-\infty$ to 
$-1$) which will be denoted by  $({\cal L})_+$ and $({\cal L})_-$,
respectively.

By using the identity 
\begin{equation}
[D,\dab ]^2 = \pa^2 \ \ ,
\label{ddb2}
\end{equation}
one 
immediately verifies that 
\begin{equation}
   [D,\dab ]^ {-1} = \pa^{-2} \, [D,\dab ]\ \ .
   \label{ddb-1}
   \end{equation}
   From (\ref{pa-1}) and (\ref{ddb-1}), it then follows that for all 
$\alpha , \beta \in {\bf R}$,
  \begin{equation}
     \left( \alpha \pa \, -\, \beta [D,\dab ] \right) ^{-1} \ =\ 
     \frac{1}{ \alpha ^2 -  \beta ^2 } \,
     \left(\alpha \pa ^{-1}\, +\, \beta [D,\dab ] ^{-1} \right)
     {\rm \ \ \ if \ \ \ \ } 
     \alpha \not= \pm \, \beta 
    \ \ . \label{albe}
      \end{equation}
     Since $\pa = \{  D ,  \dab \}$, the condition $\alpha \not= \pm \, \beta$ 
reflects the fact that the operators
     $D\dab$ and $\dab D$ are not invertible. 
     
\subsection{Superconformal covariance of a pseudodifferential symbol}

Since we are now dealing with (pseudodifferential) symbols rather than
(differential) operators, we have to generalize definition \ref{def1}.
In fact, the correspondence (\ref{eqdef1}) does not make sense in the
present case (because a symbol does not transform a field
into another field). However, we can postulate the transformation law
(\ref{translaw}).

\begin{definition}
   \label{def3}
   A pseudodifferential symbol ${\cal L}_{p,q}$, locally defined on $\Sigma$
   by (\ref{genpdo}), is {\em superconformally covariant} (or {\em covariant} for
   short) if it transforms (under a superconformal change of local coordinates)
   according to
   \begin{equation}
      \left( {\cal L} _{p,q} \right) ^{\prime} \ =\ 
      {\rm e}^{p ^{\prime}w+q ^{\prime}\bar w}
      \, {\cal L} _{p,q} \,
      {\rm e}^{-p w-q \bar w} \ \ ,
      \end{equation}
where $\ p,q,p ^{\prime},q ^{\prime} \in {\bf Z}/2 \ $ 
and $\ p+q,p ^{\prime}+q ^{\prime} \in {\bf Z} \ \ .$
\end{definition}

If ${\cal L} _{p,q}$  is a covariant 
symbol, then its differential part $({\cal L}
_{p,q})_+$ and its integral part $({\cal L}
_{p,q})_-$ are separately covariant. In particular, for the
differential part, definition \ref{def3} reduces to definition \ref{def1}
and the analysis of sections \ref{diffop} and
\ref{bolop} applies.

Once again superconformal covariance can be ensured locally by introducing
supercovariant derivatives in expression (\ref{genpdo}) 
\footnote{Note that $\{ \nabla
,\bar{\nabla} \}$ locally reads 
$\nabla \bar{\nabla} _{p,q} + \bar{\nabla} \nabla _{p,q}=
\pa -B \dab -\bar{B} D -p(D\bar B) -q(\dab B) + (p-q) B\bar B$ 
and that it is invertible because 
its leading term $\pa$ is invertible.}:
\begin{equation}
{\cal L}_{p,q} \ = \ \sum_{n=-\infty}^{n_{max}} \ \left(\ a_n \ + \ \alpha_n
 \nabla \
+\
\beta_n \bar{\nabla} \ +\ b_n [ \nabla ,\bar{\nabla} ] \right) \{ \nabla
,\bar{\nabla} \} ^n 
\label{genpdocov} \ \ .
\end{equation}
For later reference, we note that the nilpotency of the
supercovariant derivatives allows to obtain the covariant 
analogon of relations (\ref{ddb2})-(\ref{albe}) :
\begin{eqnarray*}
[\nabla ,\bar{\nabla} ]^2 &=& \{ \nabla ,\bar{\nabla} \} ^2 \\ 
{[\nabla,\bar{\nabla} ]}^{-1}
   &=& \{ \nabla ,\bar{\nabla} \} ^{-2} \, 
   [\nabla ,\bar{\nabla}  ] 
   \label{ddb-1cov} \\
\left( \alpha  \{ \nabla ,\bar{\nabla} \} \, -\, \beta [\nabla ,\bar{\nabla} ] \right)
^{-1} &= & \frac{1}{ \alpha ^2 -  \beta ^2 } \,
     \left(\alpha  \{ \nabla ,\bar{\nabla} \} ^{-1}\, +\, \beta [\nabla ,\bar{\nabla}  ]
     ^{-1} \right)  
     \label{albecov} \ \ \ \ 
     {\rm if \ \ } 
     \alpha \not= \pm \, \beta \ \ .
      \end{eqnarray*}

\subsection{Bol pseudodifferential symbols}

As in the differential case, we require a covariant 
pseudodifferential symbol to be globally defined on any compact SRS:

  \begin{definition}
      \label{def4}
A {\em Bol 
symbol} is a superconformally covariant pseudodifferential
symbol which depends on an unique superfield, namely a
projective connection.
\end{definition}  

Bol symbols can be determined
by using the same variational method as the one used for Bol operators. 
This leads
to the following result :
\begin{theorem}
   \label{theo2}
For each superconformal weight $(p,q) \in ({\bf Z}/2, {\bf Z}/2)$
such that $-(p+q) \in {\bf Z}^{*}$, there exists a Bol
 symbol which is unique up to a global factor. The 
latter is of order $n=-(p+q)$ and reads
\begin{equation}
L_{p,q} \, ({\cal R})  
 \ = \  \{ \nabla , \bar{\nabla} \} ^{n-1}
(q  \nabla \bar{\nabla}_{p,q}
-p \bar{\nabla} \nabla _{p,q} )
\ \ .
\label{respbol}
\end{equation}
\end{theorem}
For $n >0$, this expression reduces to the
Bol operator (\ref{resbol}).

The inverse of a Bol symbol (which exists if and only if $p \not= 0$ and $q  \not=
0$) is also a Bol symbol. In fact, it follows from expression (\ref{albecov})
that Bol operators and Bol symbols are related by 
the following inversion property:
\begin{eqnarray}
   L_{p,q} ^{\ -1} \ = \ - \frac{1}{p\, q} \, L_{-q,-p} \quad \quad
   {\rm if} \ p \not= 0 \ {\rm and}\ q  \not= 0.
\end{eqnarray}
  This relation can be used to determine explicitly  
  purely integral Bol symbols 
  in terms of the projective connection $\cal R$ by inverting Bol
  differential operators.
  
  Before discussing the singular cases $p=0$ or $q=0$, we briefly consider the
  symmetric case ($p=q$).

\paragraph{Symmetric Bol symbols}

For $p=q$, theorem \ref{theo2} states that
\begin{equation}
\label{lnsym}
L_n ^{sym} ({\cal R}) \ \equiv \ \{ \nabla , \bar{\nabla} \} ^{n-1} \, 
[ \nabla , \bar{\nabla} ] _{-\frac{n}{2},-\frac{n}{2}} \quad \quad (n
\not= 0)
\end{equation}
is a Bol symbol
with inverse  
$(L_n ^{sym})^{\ -1} \ =\  L_{-n} ^{sym}$. 
Interestingly enough, these
 properties can be generalized to the case $n=0$. The corresponding symbol as given by
 eq.(\ref{lnsym}) reads 
   $L_0 ^{sym}  \ = \ \pa ^{-1} \, [D, \dab ]$   
       so that it is a Bol symbol which coincides with its
       own inverse according to eq.(\ref{ddb2})~:~$(L_0 ^{sym})^{\ 2} \ =\ 1$.
 For $n>0$, the first symmetric Bol operators have been calculated
in \cite{gg} and have been shown to appear in the commutation relations of 
$N=2$ super
${\cal W}$-algebras
(see also equations (\ref{w2}) below).

\paragraph{(Anti-)chiral Bol symbols}

Among the Bol symbols given by expression (\ref{respbol}), 
the non-invertible ones 
correspond to the so-called chiral ($p=0$) and anti-chiral ($q=0$)
solutions~:
\begin{eqnarray}
\label{chj}
L_n^{chir} ({\cal R}) & \equiv & \nabla \{ \nabla , \bar{\nabla} \} ^{n-1} 
\bar{\nabla} _{0,-n}
\ = \  D
\{ \nabla , \bar{\nabla} \} ^{n-1} \dab
\\
L_n^{anti} ({\cal R}) & \equiv & \bar{\nabla} \{ \nabla , \bar{\nabla} \} ^{n-1}
\nabla _{-n,0}
\ = \ \dab
\{ \nabla , \bar{\nabla} \} ^{n-1} D
\ \ .
\nonumber
\end{eqnarray}
Although they are not invertible (as stated above), they nevertheless are related
by a kind of inversion relation, which we will now discuss.
We first note that expressions
(\ref{chj}) allow us to define the following equivalence classes of symbols, 
\begin{equation}
   \label{equivclass}
\begin{array}{llll}
   D J_n &\equiv & D \{ \nabla , \bar{\nabla} \} ^n &
   {\rm modulo} \ \ D \lambda \dab \\
   \dab K_n &\equiv & \dab \{ \nabla , \bar{\nabla} \} ^n
      &{\rm modulo} \ \ \dab \mu D \ \ \ (n \in {\bf Z}),
   \end{array}
\end{equation}
where $\lambda$ and $\mu$ are arbitrary pseudodifferential symbols. 
In the differential case $(n \geq 0)$, this amounts to restricting the
domains of definition of the operators $D J_n$ and $\dab K_n$ to 
antichiral and chiral superfields, respectively \cite{gg}.
Obviously the representatives of the equivalence classes (\ref{equivclass})
are related to the Bol
symbols (\ref{chj}) by
\begin{eqnarray}
L_n^{chir} &=& D J_{n-1} \dab \nonumber \\ 
L_n^{anti} &=& \dab K_{n-1} D \nonumber
\end{eqnarray}
and their interest consists of the fact that they satisfy 
the two (equivalent) inversion relations 
\begin{equation}
   \label{invca}
\begin{array}{lll}
   (D J_n ) ^{-1} &=& \dab K_{-n-1}\\
    (\dab K_n) ^{-1} &=& D J_{-n-1} \ \ .
   \end{array}
   \end{equation}
 The latter can be used to determine in a simple way the purely pseudodifferential 
 chiral or antichiral Bol symbols by starting from the differential
 ones. The simplest examples of symbols given by eqs. (\ref{equivclass}) read
 \begin{equation}
    \label{exbolmir}
    \begin{array}{llllll}
\dab K_{-1} &= &\dab \pa ^{-1} &
D J_{-1} &=& D \pa ^{-1} \\
\dab K_0 &=& \dab &
D J_0 &=& D \\
\dab K_1 & = & \dab [ \pa + \rr ] &
DJ_1 &=& D [ \pa - \rr ] \\
\dab K_2 & = & \dab [ \pa ^2
+3  \rr \pa + ( \dab  D \rr ) + 2 ( D \dab \rr )  + 2
\rr^2 ]
\\
    \end{array}
    \end{equation}
 and one can explicitly verify that they satisfy the relations (\ref{invca}). 
 For later reference, we note that the leading terms 
of the generic antichiral Bol symbol read ($n \in {\bf Z}$)
\begin{equation}
\label{centvir}
\dab K_{n} D \ =\ \dab \left( \pa ^n \,+\, c^{(n)} {\cal R}\pa ^{n-1}\,+\, ...
\right)D \ \ \ \ \ \ {\rm with}\ \ c^{(n)}=\frac{n(n+1)}{2}\ .
\end{equation}

\section{Covariant symbols and their applications to $N=2$ super $W$-algebras}

The antichiral Bol symbol $L_n^{anti}({\cal
    R})$ given by (\ref{chj}) is a symbol of the generic 
    form 
    \begin{equation}
\label{expmir}
\dab {\cal L}^{(n)} D \ = \ 
 \dab \left[ \pa ^n \ + \ \sum_{k=1} ^{\infty} \, a_k^{(n)}
 \pa^{n-k} \right] D \quad
(n \in {\bf Z})\ .
    \end{equation}
  It is covariant and has the property that it only depends on a projective
  connection ${\cal R}$. This suggests that, more generally,
    by requiring a generic symbol of the form (\ref{expmir})
    to be covariant and globally defined on any SRS, one
    should obtain a reparametrization of this symbol in terms of a
    projective connection and some superconformal fields of appropriate
    weight. This has been worked out in ref. \cite{gg} for differential operators
    and the extension to the pseudodifferential case is the following.\\

 Given a superconformal field ${\cal W} _{k}$ of
 weight $(k,k)$ (with $k \in {\bf N}^{*}$), the covariant
 symbol $\dab {\cal M}^{(n)}_{{\cal W}_{k}} D$ 
is defined for $n \in {\bf Z}$ by 
\begin{equation}
   \label{mw}
 \dab M^{(n)}_{{\cal W}_{k}} D \ = \ \nabar \sum_{l=0}^{\infty}
 \left\{\
A^{(n)}_{k,l}\ [(\nab \nabar)^l  {\cal W}_{k}] \; +
\; B^{(n)}_{k,l}\ [(\nabar \nab)^l {{\cal W}_{k}}]\ \right\}
\{ \nabla , \bar{\nabla} \} ^{n-k-l} \nabla _{-n-1,0} \ \ .
   \end{equation}
 It depends linearly on
  ${\cal W}_k$ and it depends on the projective connection ${\cal R}$ given by the
  Miura transformation (\ref{miu}) provided the coefficients are chosen to
  be \footnote{The binomial coefficients 
  $\left( \begin{array}{c} n \\ p
  \end{array} \right)$ are extended to $n \in {\bf Z}$ by 
  $\left( \begin{array}{c} n \\ p
  \end{array} \right) = \ds \frac{n(n-1)...(n-p+1)}{p!}$ if $p \in {\bf N}
  ^{*}$ and $\left( \begin{array}{c} n \\ p
  \end{array} \right)=1$ if $p=0$.}
  \begin{equation}
\label{coefab}
\begin{array}{ccc}
A^{(n)}_{k,l}\ =\ \frac
{\left( \begin{array}{c} n-k \\ l \end{array} \right)
\left( \begin{array}{c} k+l \\ l \end{array} \right) }
{\left( \begin{array}{c} 2k+l \\ l \end{array} \right)}
&,&
B^{(n)}_{k,l}\ =\ \frac
{\left( \begin{array}{c} n-k \\ l \end{array} \right)
\left( \begin{array}{c} k+l-1 \\ l \end{array} \right) }
{\left( \begin{array}{c} 2k+l \\ l \end{array} \right)}
\end{array}
\end{equation}
  for $l=1,..,n-k$ and $A^{(n)}_{k0}\ +\ B^{(n)}_{k0}\ =\ 1$.
 
 Instead of using the `natural'
coefficients $a_1^{(n)}, \, a_2^{(n)},
\, a_3^{(n)},...$ of expression (\ref{expmir}) 
to span the phase space of superfields, we can 
use symbols of the form (\ref{mw}) to obtain a parametrization 
which relies on a projective connection ${\cal R}$
 and of 
superconformal fields.
   \begin{theorem}
  \label{theo3} 
The most general symbol of the form 
 (\ref{expmir})
 which is 
 covariant and globally defined on any SRS is parametrized
 by a projective connection ${\cal R}$ and an infinity of superconformal
 fields  ${\cal W} _k$ of weight $(k,k)$ -- one for each value of 
 $k \in \{2,..,\infty \}$ --  according to
 \begin{equation}
\label{expcovmir}
\dab {\cal K}^{(n)} D\ =\  \bar{D} \left[
K_n + \sum^{\infty}_{k=2}
M^{(n)}_{{\cal W}_{k}} \right] D \qquad (n \in {\bf Z}) \ \ .
\end{equation}
 Here $\bar{D} K_n D$ is the antichiral Bol symbol (\ref{chj}) and 
$\bar{D} M^{(n)}_{{\cal W}_{k}}D$ is given by eq. (\ref{mw}).
\end{theorem}

The superfields 
${\cal R}$
 and ${\cal W}_{2},\, {\cal W}_{3},\,...$  
are related to the former by invertible differential
polynomials which can be explicitly determined by identifying the expressions 
(\ref{expmir}) and (\ref{expcovmir}) :
\begin{eqnarray}
a_1^{(n)} &=& c^{(n)} {\cal R} \\
\label{difpol}
a_i^{(n)} &=& \sum_{k=1}^{i} \left[
A_{k,i-k}^{(n)} (D\dab )^{i-k} {\cal W} _k \right]\, +\, 
\left[ B_{k,i-k}^{(n)} (\dab D)^{i-k} {\cal W} _k \right] \, +\, 
{\rm nonlinear \ terms}. \label{difpol2}
\end{eqnarray}
The factor $c^{(n)}$ was given in eq. (\ref{centvir}) and we have used the notation  
${\cal W} _1 \equiv a_1^{(n)}$ in the last relation.
 
 Thus we have achieved a {\em superconformal} (or {\em primary})
parametrization of the symbol (\ref{expmir}), reflecting its superconformal
covariance property.
 Of course, by starting from the chiral rather than the anti-chiral Bol symbols, one
 can repeat the whole analysis in order to achieve an analogous
 parametrization for the symbols of the form $D {\cal J}^{(n)}
 \dab$.  

\paragraph{$N=2$ super ${\cal W}_{KP}^{(n)}$-algebras}
 
We denote by ${\cal M}_n$ the manifold of covariant symbols 
of the form (\ref{expmir})
and, from now on, we restrict   our study to the case 
$n \geq 1$ for which the Bol symbol $\dab K_n D$ is purely differential 
and depends on the projective connection ${\cal R}$.
Following ref.\cite{dg}, one introduces the residue and the trace of 
a symbol $L \in {\cal M}_n$ by
\begin{equation}
{\rm res} \, L \ = \ a_{n+1}^{(n)} \ \ \ \ \ , \ \ \ \ \ 
{\rm Tr} \, L \ = \ \ds \oint d^3\uz \  {\rm res} \, L \ \,
\end{equation}
where $d^3 \uz = dzd\th d\tb$.
This trace can be used to define the pairing 
of two symbols by $\langle A,B \rangle \,=\, {\rm Tr} (AB)$. 
Let $T_L({\cal M}_n)$ and $T_L^* ({\cal M} _n)$ denote the tangent  
and cotangent
spaces of ${\cal M} _n$ at the point $L$ and 
$\Phi_U (L)$ and $\bar{\Phi} _U (L)$ denote, respectively, the chiral and 
antichiral parts of the trace ${\rm Tr} ({\rm res} [L,U])$.
Then the map
\begin{equation}
\begin{array}{llccl}
J_L &: & 
T_L^* ({\cal M} _n) &\rightarrow & T_L({\cal M}_n) \\
&&U &\mapsto & 
\ (LU)_+\,L \, -\, L\,(UL)_+ \, 
+\, L\,\Phi _U(L) \, +\, \bar{\Phi} _U(L)\, L
\label{hamap}
\end{array}
\end{equation}
determines a hamiltonian structure on the manifold ${\cal M}_n$ \cite{dg}
\footnote{The map (\ref{hamap}) has been 
determined in ref.\cite{dg} for the case of chiral operators of the form 
$D {\cal L} \dab$
and is easily transposed to the antichiral case. 
As pointed out in this reference, 
there exists a second quadratic 
hamiltonian map in the $N=2$ supersymmetric case.}.
This 
hamiltonian map is an $N=2$ extension of the usual Adler map 
\cite{adler,dick}; it defines a Poisson algebra, 
 namely the $N=2$ super ${\cal W}^{(n+1)}_{KP}$-algebra.\\
  
The superfields $ a_1^{(n)}$ and ${\cal W} _i$ ($i \geq 2$) constitute 
a {\em primary basis} for this superalgebra. In this basis, 
the Poisson brackets take a particular 
form which reflects the superconformal covariance 
property of these superfields :
\begin{eqnarray}
\left\{ a_1^{(n)}(\uz_2), a_1^{(n)}(\uz_1) \right\} &=& c^{(n)} \, {\cal
L}_2^{sym}(\rr) \,  \delta ^{(3)} (\uz _2,\uz _1)
\nonumber \\
\left\{ a_1^{(n)}(\uz_2) ,{\cal W}_k(\uz_1) \right\} &=& \left( k \, \vv \pa
-(\dab \vv )D -(D\vv )\dab +(\pa \vv ) \right)
\delta ^{(3)} (\uz _2,\uz _1)
\label{w2}
\\
\left\{ {\cal W}_k (\uz_2) ,{\cal W}_l(\uz_1) \right\} &=&
\left(c_{kl} ^{(n)}\, {\cal
L}_{k+l}^{sym}(\rr) \, +\, ... \right) \delta
^{(3)} (\uz _2,\uz _1) \nonumber
\end{eqnarray}
The first relation represents the $N=2$ Virasoro superalgebra. 
The differential operator $L_2^{sym} ({\cal R})$ is the symmetric Bol operator given 
by (\ref{lnsym}) and explicitly reads $L_2^{sym} ({\cal R})=
\pa [D,\dab ]
+ {\cal R} \pa
- (D{\cal R} ) \dab
- (\dab {\cal R} ) D
+ (\pa {\cal R})$ ; hence the Virasoro central term 
reads $\frac{1}{2} n(n+1)\,\pa [D,\dab ] \delta
^{(3)} (\uz _2,\uz _1)$.
The second relation reflects the fact that ${\cal W}_k$ is a
superconformal field of weight $(k,k)$. In the last relation, we have only written the
terms which do not depend on the primary fields ${\cal W}_i \ \ (i \geq 2)$.
Since 
the leading term of the operator ${\cal
L}_{k+l}^{sym}(\rr)=\pa ^{k+l-1} [D,\dab ] \, +\, ...$ does not depends on any 
field, the central term reads 
$c_{kl} ^{(n)} \pa ^{k+l-1} [D,\dab ] \delta
^{(3)} (\uz _2,\uz _1)$.

\paragraph{$N=2$ super ${\cal W}_n$-algebras}
 
Interestingly enough, the primary parametrization (\ref{expcovmir}) allows us to
split automatically the differential and integral parts of the symbol 
$\dab {\cal K}^{(n)} D$. In fact, if
$k>n$, the symbol (\ref{mw}) is purely integral: $\dab M^{(n)}_{{\cal W}_{k}}
D = (\dab M^{(n)}_{{\cal W}_{k}} D)_-$. On the contrary, if $k\leq n$, 
it is a differential operator  (the summation over $l$ is going from $0$ to
$n-k$ since the coefficients (\ref{coefab}) vanish for larger values of
$l$): $\dab M^{(n)}_{{\cal W}_{k}} D = (\dab M^{(n)}_{{\cal W}_{k}} D)_+$. 
 
Thus a symbol $L \in {\cal M}_n$ can be easily divided 
into its differential and integral parts :
\begin{equation}
\label{dec+-}
L_+ \ =\  \bar{D} \left[
K_n + \sum^{n}_{k=2}
M^{(n)}_{{\cal W}_{k}} \right] D \ \ \ \ \ ,\ \ \ \ \ 
L_- \ = \ \dab \sum_{k=n+1}^{\infty}
M^{(n)}_{{\cal W}_{k}} D \ \ \ .
\end{equation}
According to eqs.(\ref{hamap}), $L_-=0$ implies $J_L(U)_-=0$ for all $U$. 
Thus, one can impose the constraint $L_-=0$. Hence 
the superfields ${\cal W}_k$ with $k > n$ which parametrize $L_-$ 
generate an ideal ${\cal I}_n$ of 
${\cal W}_{KP}^{(n+1)}$ \cite{fmrtopo}. This ideal is 
centerless so that the central charges $c_{kl}^{(n)}$ in (\ref{w2}) 
vanish if both $k> n$ and $l>n$.
 
Moreover, the constraint $L_-=0$ allows for a hamiltonian reduction : 
the quotient of ${\cal W}_{KP}^{(n+1)}$ 
by its ideal ${\cal I}_n$ is isomorphic to 
the $N=2$ super ${\cal W}_{n+1}$-algebra generated by the 
differential operator $L_+$. 
According to eq.(\ref{expcovmir}), the latter is parametrized by the 
$n$ superfields
${\cal R}$ and ${\cal W}_{2},..,{\cal W}_{n}$ which thus span 
a primary basis of the $N=2$ super ${\cal W}_{n+1}$-algebra  
as shown in ref. \cite{gg}.

\section{Conclusion}

By studying $N=2$ covariant symbols, we 
have achieved a classification of the Bol symbols 
which are characterized by their dependence on a superprojective connection.
Among them, the antichiral Bol symbols $\dab K_n D$ are 
of particular interest since they 
are a special case of symbols $\dab {\cal L}^{(n)} D$ 
of the form (\ref{expmir}) 
whose manifold can be endowed with a hamiltonian structure leading to 
the $N=2$ super ${\cal W}_{KP}^{(n+1)}$-algebra.  
We have parametrized such symbols by using a superprojective connection 
and an infinity of primary superfields ${\cal W}_k$ of 
weights $k=2,..,\infty$. This provides 
us with a primary basis of 
generators for the $N=2$ super ${\cal W}_{KP}^{(n+1)}$-algebra  
and, by reduction, also one for the super ${\cal W}_{n+1}$-algebra.
If expressed in this basis, the Poisson brackets take a  
form which reflects 
their $N=2$ superconformal symmetry. 
This allowed us to study the central terms: the 
Virasoro central charge $c^{(n)}$ of 
${\cal W}_{KP}^{(n+1)}$ can be explicitly determined. In principle, the 
other charges $c_{kl} ^{(n)}$ could also be computed by inverting 
relation (\ref{difpol2}) and proceeding along the lines of ref. \cite{dfiz.1}.
We conjecture that, 
due to the choice of the primary basis, 
these central charges turn out to be diagonal ($c_{kl} ^{(n)} = c_{kl} ^{(n)}
\delta _{kl}$)  
as in the nonsupersymmetric case.

\paragraph{Acknowledgements}
 The author would like to thank Fran\c{c}ois Gieres for his careful reading of the
 manuscript and his suggestions.

\nocite{*} 
\bibliographystyle{unsrt}
\bibliography{article}

\end{document}